\documentclass[aps,prfluids,reprint,amsmath,amssymb,longbibliography,superscriptaddress]{revtex4-2}

\usepackage[T1]{fontenc}
\usepackage{bm,graphicx,booktabs,mathtools,amsthm}
\usepackage{hyperref}
\hypersetup{colorlinks=true, linkcolor=blue, citecolor=blue, urlcolor=blue}

\newcommand{\grad}{\nabla}
\newcommand{\Div}{\nabla\!\cdot}
\newcommand{\dd}{\mathrm{d}}

\newcommand{\ve}{\bm{v}}

\theoremstyle{definition}
\newtheorem{definition}{Definition}
\theoremstyle{plain}

\newtheorem{theorem}{Theorem}

\begin{document}

\title{A mixed-metric two-field framework for turbulence:\\
Emergent stress anisotropy and wall asymptotics from a single scalar}

\author{Marcial Sanchis-Agudo}
\email{sanchis@kth.se}
\affiliation{FLOW, Engineering Mechanics, KTH Royal Institute of Technology, SE-100 44 Stockholm, Sweden}

\author{Ricardo Vinuesa}
\affiliation{Department of Aerospace Engineering, University of Michigan, Ann Arbor, MI 48109, USA}
\affiliation{FLOW, Engineering Mechanics, KTH Royal Institute of Technology, SE-100 44 Stockholm, Sweden}

\date{\today}

\begin{abstract}
In our previous work~\cite{SanchisAgudoVinuesa2025PRL}, we argued that viscous dissipation in turbulence can be understood as the macroscopic imprint of microscopic path uncertainty, and showed that a kernel variance field $s(y)$ constrained by a balance condition yields both the Kolmogorov scales and the logarithmic law of the wall from a single stochastic principle. In the present work we promote $s$ to a dynamical field $s(\bm{x},t)$ with units of kinematic viscosity and develop a two-field framework in which the velocity $\ve$ and an \emph{intermittency} (or stochastic diffusivity) field $s$ evolve in a coupled way. The effective viscosity is $\nu_{\mathrm{eff}}=\nu_0+s$, but the stress tensor is generalized to include a non-linear closure driven by the commutator of strain and rotation, $[\bm{S}, \bm{\Omega}]$, capturing emergent anisotropy. The evolution of $s$ is defined as a mixed-metric gradient flow: a Wasserstein-2 gradient flow for morphology, $\Div(s\grad s)$, combined with a local $L^2$ gradient flow driven by an objective coupling term $q$. The coupling is decomposed as $q=q_{\mathrm{prod}}-q_{\mathrm{relax}}$, where production is driven by a vortex-stretching invariant, $\mathcal{I} = \|\bm{S}\boldsymbol{\omega}\|^2$. This choice ensures that production vanishes identically in strictly two-dimensional flows. We show that, under standard assumptions of constant stress, high Reynolds number and overlap-layer scale invariance, the only scale-invariant overlap-layer solution of the mixed-metric equation is $s(y)\propto y$, which recovers the logarithmic velocity profile. Thus the same mixed-metric equation organizes both wall-resolved and wall-modeled asymptotics within a single, energetically constrained framework. We conclude by interpreting the framework as the equilibrium (Markovian) limit of a generalized viscoelastic theory built on the geometry of uncertainty introduced in our previous work.
\end{abstract}

\maketitle

\section{Introduction}
\label{sec:intro}

A significant fraction of the challenge in predicting turbulent flows is related to the behavior of the near-wall region. In our previous work~\cite{SanchisAgudoVinuesa2025PRL} we took a geometric viewpoint and asked: what if viscous dissipation is the macroscopic imprint of a crowd trying to cross a plaza? Imagine crossing a ``grand plaza'' from A to B. In the empty plaza you walk along a straight geodesic, the world of least action: this is Arnold's ideal fluid~\cite{Arnold1966}. In a crowded plaza, your \emph{intention} (mean path) is still approximately geodesic, but your actual trajectory jitters as you dodge others. You travel through a cloud of paths, and the trip takes longer on average. In our previous work~\cite{SanchisAgudoVinuesa2025PRL} we encoded that cloud by a kernel variance field and showed that constraining this exploratory field near a wall is enough to recover two pillars of turbulence theory: the Kolmogorov scales (velocity, time and length) and the logarithmic law of the wall.

The present paper asks a sharper question: \emph{what is $s$ as a physical field}, and can we write an evolution equation for it that is both constrained by geometry and useful for modeling?

\vspace{0.5em}
\noindent\textbf{From algebraic closures to a dynamical $s$-field.}

Classical models treat unresolved turbulence through an effective viscosity, whether as an algebraic eddy viscosity
\[
\nu_t = (C_S \Delta)^2 K
\quad\text{or}\quad
\nu_t \sim \ell_m^2 |U'|
\]
in LES and mixing-length theories~\cite{Smagorinsky1963,Lilly1967,Schumann1975,Pope2000}, or through more elaborate two-equation closures such as $k$--$\epsilon$ or $k$--$\omega$~\cite{LaunderSpalding1974,Menter1994}. In all cases, the effective viscosity is built as a functional of the resolved velocity gradient and a few extra scalars, chosen to satisfy symmetries and match data, but not derived from an underlying evolution principle.

Our previous work~\cite{SanchisAgudoVinuesa2025PRL} suggested a different starting point: viscosity as a consequence of microscopic path uncertainty. In that picture, $s$ is not just a tuning function but the field that quantifies how much trajectories \emph{explore} around the mean flow. Here we adopt that interpretation and promote $s$ to a space-time field $s(\bm{x},t)$ that co-evolves with the velocity:
\begin{itemize}
  \item the velocity $\ve$ describes the \emph{exploit} dynamics, i.e.\ the organized mean motion across the plaza;
  \item the scalar field $s$ describes the \emph{explore} dynamics, i.e.\ the local intensity of microscopic wandering that gives rise to dissipation.
\end{itemize}
Throughout this paper we assign $s$ the units of kinematic viscosity ($L^2/T$) and interpret it as a \emph{stochastic diffusivity} (a variance \emph{rate} in path space). The effective viscosity is then
\begin{equation}
\nu_{\text{eff}}(\bm{x},t) = \nu_0 + s(\bm{x},t),
\label{eq:intro_nueff}
\end{equation}
with $\nu_0$ being the molecular contribution and $s$ the turbulent one.

\paragraph*{Relation to the kernel variance in our previous work.}
In our previous work~\cite{SanchisAgudoVinuesa2025PRL}, the symbol $s(y)$ appeared as a variance scale with units of length squared. Here $s(\bm{x},t)$ has units of $L^2/T$, because it enters additively in $\nu_{\text{eff}}=\nu_0+s$ and plays the role of a stochastic diffusivity. The two uses are consistent if one introduces a small kernel time scale $\Delta t_*$ and writes the variance of the underlying Brownian exploration as
\begin{equation}
\sigma^2(y) \;\sim\; 2\,\nu_{\text{eff}}(y)\,\Delta t_* \;\equiv\; 2\,[\nu_0+s(y)]\,\Delta t_*.
\end{equation}
In this coarse-grained picture, the ``variance profile'' of our previous work is $\sigma^2(y)$, whereas the present intermittency field is
\begin{equation}
s(y)=\nu_{\text{eff}}(y)-\nu_0.
\end{equation}
Up to the constant factor $2\Delta t_*$, both encode the same wall-normal structure of path uncertainty: in our previous work we emphasized the (dimensionful) variance, here we take the corresponding \emph{variance rate} and use it as a dynamical field. For this reason we reuse the symbol $s$ in the two-field framework, implicitly absorbing $\Delta t_*$ into its definition.

The central question of this paper is how to define the evolution of $s$ in a way that is:
\begin{enumerate}
  \item variationally structured, so that some part of its dynamics is fixed by a gradient-flow principle rather than ad hoc;
  \item objectively coupled to the resolved velocity gradient, through an invariant that detects genuinely three-dimensional stretching;
  \item compatible with wall physics, in the sense that it recovers both the viscous sublayer and the logarithmic overlap layer as asymptotic limits.
\end{enumerate}

\vspace{0.5em}
\noindent\textbf{Summary of the framework.}

We propose a two-field framework in which:
\begin{itemize}
  \item the momentum equation is closed by $\nu_{\text{eff}} = \nu_0 + s$ plus an emergent non-linear stress term that accounts for rotational anisotropy,
  \item the field $s$ evolves according to a \emph{mixed-metric gradient flow}: a Wasserstein-2 gradient flow of a simple ``morphology'' energy, yielding the degenerate diffusion $\Div(s\grad s)$ familiar from porous-medium theory~\cite{JKO1998,Otto2001,Vazquez2007PME}, plus a local $L^2$ gradient flow of a reaction potential that encodes production and relaxation.
\end{itemize}
We are explicit that this is a \emph{modeling framework structured by axioms}, not a full derivation from the Schr\"odinger-Bridge formalism of our previous work~\cite{SanchisAgudoVinuesa2025PRL}: the SB perspective motivates the existence and additivity of a scalar diffusivity field, but the concrete choices for the evolution of $s$ are specified below as physically guided modeling assumptions.

A key structural element is the invariant driving production. We argue that the natural choice is the squared norm of the vortex-stretching vector,
\begin{equation}
\mathcal{I}(\bm{S},\boldsymbol{\omega}) = \|\bm{S}\boldsymbol{\omega}\|^2,
\end{equation}
where $\bm{S}$ is the strain-rate tensor and $\boldsymbol{\omega}=\nabla\times\bm{v}$ the vorticity. This quantity measures how strongly the flow stretches and tilts vortical structures, vanishes identically in strictly two-dimensional flows, and has the correct dimension to enter the source term of the $s$-equation.

The rest of the paper is organized as follows. In Sec.~\ref{sec:framework} we formalize the two-field setup and notation. In Sec.~\ref{sec:mixedflow} we derive the mixed-metric gradient flow for $s$ and specify the production and relaxation terms. In Sec.~\ref{sec:asymptotics} we analyze the wall asymptotics and show that the same equation yields $s\sim y^2$ as $y\to 0$ and $s\propto y$ in the overlap. Sec.~\ref{sec:thermostat} interprets the production coefficient as a Kolmogorov-based thermostat. In Sec.~\ref{sec:discussion} we provide a physical interpretation of the geometry of uncertainty, before concluding in Sec.~\ref{sec:conclusion}.

\section{Two-Field Framework}
\label{sec:framework}

We consider an incompressible fluid described by a velocity field $\ve(\bm{x},t)$ and a second scalar field $s(\bm{x},t)$, which we interpret as the local \emph{intermittency} or stochastic diffusivity associated with the microscopic path-uncertainty process introduced in our previous work~\cite{SanchisAgudoVinuesa2025PRL}.

\paragraph*{Kinematics and invariants.}
The velocity gradient is decomposed into symmetric and antisymmetric parts,
\begin{equation}
\grad\ve = \bm{S} + \bm{\Omega},
\end{equation}
with strain-rate tensor $\bm{S}$ and rotation-rate tensor $\bm{\Omega}$ given by
\begin{align}
S_{ij} &= \frac{1}{2}\left(\frac{\partial v_i}{\partial x_j}
+ \frac{\partial v_j}{\partial x_i}\right),\\
\Omega_{ij} &= \frac{1}{2}\left(\frac{\partial v_i}{\partial x_j}
- \frac{\partial v_j}{\partial x_i}\right).
\end{align}
The vorticity is $\boldsymbol{\omega} = \grad\times\ve$, and the scalar shear magnitude is
\begin{equation}
K := \sqrt{2\,\bm{S}:\bm{S}}.
\end{equation}

\paragraph*{Momentum equation with emergent anisotropy.}
To close the momentum equation, we posit that the effective stress tensor is determined by the scalar magnitude $s$ but possesses a structure dictated by the local flow geometry. We retain the incompressible Navier--Stokes form for the left-hand side and interpret the new stress as the macroscopic footprint of the microscopic path uncertainty:
\begin{equation}
\partial_t\ve + (\ve\!\cdot\!\grad)\ve
= -\grad p + \Div\!\bm{\Sigma},
\qquad \Div \ve=0,
\label{eq:momentum}
\end{equation}
where the total viscous stress tensor $\bm{\Sigma}$ is defined as:
\begin{equation}
\bm{\Sigma} = 2(\nu_0 + s)\bm{S} + \bm{\tau}_{\text{aniso}}.
\end{equation}
The term $2s\bm{S}$ represents an eddy-viscosity (isotropic) contribution to the deviatoric stress, consistent with identifying $\nu_{\text{eff}}=\nu_0+s$ as the sum of molecular and turbulent (stochastic) diffusivities. The term $\bm{\tau}_{\text{aniso}}$ captures stress anisotropy that emerges from the interaction between strain and rotation. Motivated by explicit algebraic Reynolds-stress modeling ideas, we define:
\begin{equation}
\bm{\tau}_{\text{aniso}} = C_2 \frac{s}{K} \left( \bm{S}\bm{\Omega} - \bm{\Omega}\bm{S} \right).
\label{eq:aniso_closure}
\end{equation}
Here, the commutator $[\bm{S}, \bm{\Omega}] = \bm{S}\bm{\Omega} - \bm{\Omega}\bm{S}$ quantifies the misalignment of principal stress and strain axes due to mean rotation. The coefficient $s/K$ ensures dimensional consistency ($L^2/T^2$) while linking the intensity of the anisotropy to the uncertainty field $s$. In numerical implementation, the factor $1/K$ requires regularization (e.g., $1/\max(K, \epsilon)$) to avoid singularities in weak-strain regions.

We stress that Eq.~\eqref{eq:aniso_closure} is a modeling choice (not derived from the Schr\"odinger-Bridge formalism) guided by objectivity and by the requirement that the anisotropic contribution vanish when $s=0$.

The central question is how to define the evolution of $s$ in a principled way.

\section{Mixed-Metric Gradient Flow for $s$}
\label{sec:mixedflow}

We consider the evolution of $s$ as a superposition of two gradient flows of an energy functional $E[s]$, taken in different metrics.

\begin{definition}[Mixed-metric gradient flow]
\label{def:mixedflow}
Let
\begin{align}
E[s] &= E_{\mathrm{morph}}[s] + E_{\mathrm{react}}[s],
\\
E_{\mathrm{morph}} &= \int \frac{1}{2}s^2\,\dd\bm{x},\\
E_{\mathrm{react}} &= \int V(s;\bm{S},\boldsymbol{\omega})\,\dd\bm{x}.
\end{align}
The evolution of $s$ is defined as the sum of:
\begin{itemize}
  \item the Wasserstein-2 gradient flow of $E_{\mathrm{morph}}$ (morphology),
  \item the $L^2$ gradient flow of $E_{\mathrm{react}}$ (reaction/coupling),
\end{itemize}
advected by the resolved velocity $\ve$.
\end{definition}

\paragraph*{A clean intermediate step.}
Define
\begin{equation}
u := \tfrac{1}{2}s^2 \qquad (u\ge 0).
\label{eq:u_def}
\end{equation}
Note that
\begin{equation}
\Delta u \;=\; \Div(s\grad s).
\end{equation}
In particular, the Wasserstein-2 gradient flow of $E_{\mathrm{morph}}[s]=\int \tfrac{1}{2}s^2\,\dd\bm{x}$ yields
\begin{equation}
\big(\partial_t s\big)_{\mathrm{morph}}
=
\Div\!\left(s\,\grad\frac{\delta E_{\mathrm{morph}}}{\delta s}\right)
=
\Div(s\grad s)
=
\Delta u,
\end{equation}
which makes explicit the porous-medium-type (degenerate) structure in the $s$-variable.

Applying the mixed-metric construction leads to
\begin{equation}
\partial_t s + \ve\!\cdot\!\grad s
=
\underbrace{\Div(s\grad s)}_{\text{morphology}}
+
\underbrace{q(\ve,s)}_{\text{coupling}},
\label{eq:s_evolution}
\end{equation}
where the coupling term $q$ encodes exchange with the resolved flow through $V$.

\paragraph*{Steady balance and Poisson resolvent.}
In steady pockets where both the time derivative and advection are negligible relative to morphology and coupling, the equation reduces to
\begin{equation}
\Delta u = -q,
\label{eq:Poisson_balance}
\end{equation}
with formal solution
\begin{equation}
u = (-\Delta)^{-1} q = \int_0^\infty e^{t\Delta} q\,\dd t,
\end{equation}
i.e.\ $u$ is obtained by convolving $q$ with the heat kernel. This resolvent structure mirrors the way the Dirichlet kernel organized the wall layer in our previous work~\cite{SanchisAgudoVinuesa2025PRL}.

\subsection{Coupling term and dimensional consistency}
\label{sec:coupling}

We decompose
\begin{equation}
q = q_{\mathrm{prod}} - q_{\mathrm{relax}},
\end{equation}
where $q_{\mathrm{prod}}$ is a production term driven by three-dimensional vortex stretching, and $q_{\mathrm{relax}}$ represents relaxation driven by resolved shear.

We require $q$ to have dimension
\begin{equation}
[q] = \frac{L^2}{T^2},
\end{equation}
so that all terms in Eq.~\eqref{eq:s_evolution} have consistent units. The basic quantities are summarized in Table~\ref{tab:dimensions}.

\begin{table}[t]
\caption{Dimensional consistency of quantities entering the $s$-equation.}
\centering
\begin{tabular}{l c c}
\toprule
Quantity & Symbol & Dimension \\
\midrule
Intermittency (stochastic diffusivity) & $s$ & $L^2/T$ \\
Shear magnitude & $K$ & $T^{-1}$ \\
Strain-rate tensor & $\bm{S}$ & $T^{-1}$ \\
Vorticity & $\boldsymbol{\omega}$ & $T^{-1}$ \\
Production invariant & $\mathcal{I} = \|\bm{S}\boldsymbol{\omega}\|^2$ & $T^{-4}$ \\
\midrule
Linear relaxation & $sK$ & $L^2/T^2$ \\
Quadratic relaxation (dimensionless $C_4$) & $(C_4/\nu_0)\, s^2 K$ & $L^2/T^2$ \\
Production source & $C_I (\nu_0+s)\mathcal{I}$ & $L^2/T^2$ \\
\bottomrule
\end{tabular}
\label{tab:dimensions}
\end{table}

For clarity (and to keep free parameters dimensionless), we write the saturation coefficient as $(C_4/\nu_0)$ with $C_4$ dimensionless, and we parametrize the production coefficient as $C_I=C_k\tau_\eta^3$ with $C_k$ dimensionless (see Sec.~\ref{sec:thermostat}).

\subsection{Production invariant: vortex stretching and causal closure}
\label{sec:prod_invariant}

A central modeling choice is the invariant driving production. We impose four requirements:
\begin{enumerate}
  \item \emph{Objectivity:} it must be frame-indifferent, i.e.\ unaffected by superposed rigid-body motions~\cite{TruesdellNoll1965,Speziale1987}.
  \item \emph{Dimensionality:} it must distinguish strictly two-dimensional limits (no vortex stretching) from genuinely three-dimensional turbulence~\cite{Kraichnan1967}.
  \item \emph{Consistency:} it must have dimension $T^{-4}$ (Table~\ref{tab:dimensions}).
  \item \emph{Causality:} it must be tied to vortex stretching/tilting, the kinematic mechanism behind the forward cascade~\cite{TennekesLumley1972,Pope2000}.
\end{enumerate}

\paragraph*{Kinematic requirement.}
Introduce the vorticity $\boldsymbol{\omega} = \grad\times\bm{v}$ and the vortex-stretching vector $\bm{w} := \bm{S}\,\boldsymbol{\omega}$.
Guided by the vorticity-transport equation, we adopt:

\begin{description}
\item[Axiom 1 (Causal link).]
The local production rate $q_{\mathrm{prod}}$ must be proportional to the square of the vortex-stretching rate,
\begin{equation}
q_{\mathrm{prod}} \;\propto\; \text{(flow intensity)} \times \|\bm{S}\boldsymbol{\omega}\|^2,
\end{equation}
so that $q_{\mathrm{prod}}\ge 0$ and production vanishes whenever there is no stretching or tilting of vorticity.
\end{description}

We choose
\begin{equation}
\mathcal{I}(\bm{S},\boldsymbol{\omega})
=
\|\bm{S}\boldsymbol{\omega}\|^2
=
(\bm{S}\boldsymbol{\omega})\cdot(\bm{S}\boldsymbol{\omega}),
\qquad
\boldsymbol{\omega} = \grad\times\bm{v}.
\label{eq:I_def}
\end{equation}

In any strictly two-dimensional flow $\bm{v}(x,y,t)=(u(x,y,t),v(x,y,t),0)$ with no dependence on the third coordinate, the vorticity is purely out-of-plane and the symmetric strain cannot stretch it; one finds $\bm{S}\boldsymbol{\omega}=\bm{0}$ and therefore
\begin{equation}
\mathcal{I}(\bm{S},\boldsymbol{\omega}) = 0
\quad\text{for all strictly two-dimensional flows.}
\end{equation}

\paragraph*{Dimensional closure.}
We model the source as
\begin{equation}
q_{\mathrm{prod}} = C_I(\nu_0,s,\epsilon)\,(\nu_0+s)\,\mathcal{I}(\bm{S},\boldsymbol{\omega}),
\label{eq:q_prod}
\end{equation}
so $(\nu_0+s)\mathcal{I}$ has dimension $L^2/T^{5}$ and therefore $[C_I]=T^3$.

Our previous work~\cite{SanchisAgudoVinuesa2025PRL} motivates tying $C_I$ to the Kolmogorov time scale
\begin{equation}
\tau_\eta = \left(\frac{\nu_0}{\epsilon}\right)^{1/2}.
\end{equation}

\begin{description}
\item[Axiom 2 (Dimensional closure).]
The coefficient $C_I$ is set by the local dissipation time scale, i.e.\ we take
\begin{equation}
C_I(\bm{x},t) = C_k\,\tau_\eta(\bm{x},t)^3,
\end{equation}
with $C_k$ dimensionless.
\end{description}

\subsection{Relaxation and saturation}
\label{sec:relaxation}

Relaxation is modeled as
\begin{equation}
q_{\mathrm{relax}} = sK + \frac{C_4}{\nu_0}\, s^2 K,
\label{eq:q_relax}
\end{equation}
where $C_4$ is dimensionless. The linear term $sK$ is dominant in the scale-invariant overlap layer, while the quadratic term provides saturation in regions where $s$ becomes large.

Collecting all contributions,
\begin{equation}
q(\ve,s)
=
C_I(\nu_0,s,\epsilon)\,(\nu_0+s)\,\mathcal{I}(\bm{S},\boldsymbol{\omega})
\;-\;
sK
\;-\;
\frac{C_4}{\nu_0}\,s^2 K.
\label{eq:q_full}
\end{equation}

\section{Wall Asymptotics}
\label{sec:asymptotics}

We examine the behavior of Eq.~\eqref{eq:s_evolution} near a planar wall, focusing on two asymptotic regimes relevant in LES practice:
\begin{itemize}
  \item \emph{WRLES} (wall-resolved LES): the near-wall region is sufficiently resolved that the modeled contribution must vanish smoothly at the wall;
  \item \emph{WMLES} (wall-modeled LES): the overlap is treated via an asymptotic model consistent with constant-stress arguments.
\end{itemize}

We consider a canonical plane channel with wall-normal coordinate $y$, wall at $y=0$, and mean velocity $U(y)$ in the streamwise direction. We assume statistical stationarity and homogeneity in the streamwise and spanwise directions.

\subsection{WRLES limit: near-wall behavior}
\label{sec:WRLES}

In the viscous sublayer ($y^+\lesssim 10$), the mean flow is close to a simple shear. While instantaneous near-wall turbulence is three-dimensional, vortex stretching is strongly constrained; in the strict two-dimensional shear limit the invariant $\mathcal{I}$ vanishes identically. Thus, as a leading-order asymptotic model for the inner-most region, we take
\begin{equation}
q_{\mathrm{prod}} \approx 0,
\end{equation}
and the steady $s$-equation reduces to a balance between morphology and relaxation,
\begin{equation}
\Delta u \approx q_{\mathrm{relax}},
\qquad
u = \tfrac{1}{2}s^2.
\label{eq:Poisson_nearwall}
\end{equation}

In a one-dimensional approximation ($s=s(y)$, $K\approx K_0$ close to the wall), this reads
\begin{equation}
\frac{\dd^2}{\dd y^2}\left(\tfrac{1}{2}s^2\right)
\;\approx\;
sK_0 + \frac{C_4}{\nu_0}\,s^2 K_0.
\label{eq:Poisson_1D}
\end{equation}
Assuming no-slip and finite variance rate at the wall, we impose
\begin{equation}
s(0) = 0,
\qquad
s(y)\ge 0,
\qquad
\big(s\,\partial_y s\big)\big|_{y=0}=0,
\end{equation}
where the last condition is the no-flux morphology boundary condition.

\paragraph*{A slightly more robust leading-order argument.}
Assume a smooth expansion $s(y)=a_1 y + a_2 y^2 + \mathcal{O}(y^3)$. The no-flux condition implies $s\,s'|_{y=0}=0$, which allows $a_1$ but does not enforce it. Substituting into \eqref{eq:Poisson_1D} shows that if $a_1\neq 0$ then the left-hand side behaves like a constant while the right-hand side behaves like $\mathcal{O}(y)$, which is inconsistent. Hence $a_1=0$ and the first nonzero term is quadratic.

Equivalently, seeking a power law $s(y)\sim c\,y^n$ as $y\to 0$ yields the same conclusion. With $u=\tfrac{1}{2}s^2\sim \tfrac{1}{2}c^2 y^{2n}$ one has $u''\sim \tfrac{1}{2}c^2(2n)(2n-1)y^{2n-2}$, whereas the leading relaxation term is $sK_0\sim cK_0 y^n$. Balancing exponents gives $2n-2=n$, i.e.\ $n=2$:
\begin{equation}
s(y) \sim c\,y^2,\qquad u(y)\sim \tfrac{1}{2}c^2 y^4
\quad\text{as } y\to 0.
\label{eq:s_y2}
\end{equation}

\paragraph*{Connection to practical WRLES damping and the $y^3$ condition.}
In many wall-resolved LES implementations based on Smagorinsky-type or dynamic SGS models, the eddy viscosity is multiplied by a van-Driest damping function so that the modeled turbulent viscosity behaves as
\begin{equation}
\nu_t^{\text{SGS}}(y^+) \;\propto\; (y^+)^3
\quad\text{as } y^+\to 0,
\end{equation}
while recovering $\nu_t\propto y$ in the logarithmic region; see, for instance, H\"artel \& Kleiser~\cite{HaertelKleiser1998JFM} and van Driest~\cite{vanDriest1956}. In our framework, the intrinsic small-$y$ scaling is $s\sim y^2$, i.e.\ $\nu_t^{\text{core}}(y)=s(y)\sim y^2$ in the resolved limit.

If one wishes to enforce $\nu_t\sim y^3$ at very small $y^+$ for coarse grids, this can be achieved by composing $s$ with a damping function $D(y^+)$, for example
\begin{equation}
\nu_t^{\text{WRLES}}(y) = D(y^+)\,s(y),
\qquad D(y^+)\sim y^+ \quad (y^+\to 0),
\end{equation}
so $s\sim y^2$ together with $D\sim y^+$ yields $\nu_t^{\text{WRLES}}\sim y^3$ without modifying the core two-field PDE.

\subsection{WMLES limit: overlap layer and logarithmic law}
\label{sec:WMLES}

In the overlap layer, we assume:
\begin{enumerate}
  \item A constant-stress region: $\tau_{xy}\approx \rho u_\tau^2$.
  \item High Reynolds number: $s\gg \nu_0$, so $\nu_{\mathrm{eff}}\approx s$.
  \item Overlap-layer scale invariance: no length scale other than $y$ enters the leading-order balance~\cite{Townsend1976,Marusic2013,Hoyas2024PRF}.
\end{enumerate}

\paragraph*{Mean-momentum balance.}
In a steady, fully developed channel,
\begin{equation}
(\nu_0+s)U' \approx u_\tau^2,
\end{equation}
where $U'=\dd U/\dd y$. The anisotropic stress term \eqref{eq:aniso_closure} does not modify the mean shear-stress balance for parallel shear flow.

In the overlap limit $\nu_0\ll s$,
\begin{equation}
s(y)\,U'(y) \approx u_\tau^2.
\label{eq:sUprime_const}
\end{equation}

\paragraph*{Scale invariance selects $s(y)\propto y$.}
To determine the $y$-dependence of $s$, assume that the leading-order overlap dynamics introduce no length scale other than $y$. Consider the one-dimensional morphology operator,
\begin{equation}
\mathcal{L}[s] := \frac{\dd}{\dd y}\left(s(y)\frac{\dd}{\dd y}\right).
\end{equation}
If $s(y)\sim y^\alpha$, then under dilation $y\mapsto \lambda y$,
\begin{equation}
\mathcal{L}[s](\lambda y)\sim \lambda^{\alpha-2}\,\mathcal{L}[s](y).
\end{equation}
Overlap-layer scale invariance (no intrinsic length beyond $y$) selects $\alpha=1$, hence
\begin{equation}
s(y)=c\,y,
\label{eq:s_linear}
\end{equation}
with $c$ a constant with units of velocity.

Substituting \eqref{eq:s_linear} into \eqref{eq:sUprime_const} yields
\begin{equation}
c\,y\,U'(y) \approx u_\tau^2
\quad\Rightarrow\quad
U'(y)\approx \frac{u_\tau^2}{c}\frac{1}{y},
\end{equation}
so
\begin{equation}
U(y)\approx \frac{u_\tau^2}{c}\ln y + B.
\end{equation}
Identifying $c=\kappa u_\tau$, we recover
\begin{equation}
U^+(y^+) = \frac{1}{\kappa}\ln y^+ + B^+.
\end{equation}

\begin{theorem}[Logarithmic law from scale-invariant $s$]
\label{thm:loglaw}
Under the assumptions of constant stress, high Reynolds number ($\nu_0\ll s$), and overlap-layer scale invariance, the only nontrivial overlap-layer profile satisfies $s(y)\propto y$, yielding a logarithmic mean velocity profile.
\end{theorem}

\section{Energetic Interpretation and the $C_I$ Thermostat}
\label{sec:thermostat}

The production coefficient $C_I$ in Eq.~\eqref{eq:q_prod} is interpreted energetically. Define the local dissipation rate per unit mass by
\begin{equation}
\epsilon(\bm{x},t)=\nu_{\mathrm{eff}}K^2.
\end{equation}
In our previous work~\cite{SanchisAgudoVinuesa2025PRL} we motivated the Kolmogorov time scale
\begin{equation}
\tau_\eta = \left(\frac{\nu_0}{\epsilon}\right)^{1/2}
\end{equation}
as a dissipation-controlled microscopic adjustment time. A minimal choice consistent with $[C_I]=T^3$ is
\begin{equation}
C_I(\bm{x},t)=C_k\,\tau_\eta^3
= C_k\left(\frac{\nu_0}{\epsilon(\bm{x},t)}\right)^{3/2},
\label{eq:CI_def}
\end{equation}
with $C_k$ dimensionless. This makes production strongest where stretching is intense and dissipation time scales are long, and weaker where dissipation is already strong.

\section{Relation to Classical Turbulence Models}
\label{sec:relation}

\paragraph*{Eddy viscosity and mixing length.}
In Smagorinsky-type LES~\cite{Smagorinsky1963,Lilly1967}, one writes
\begin{equation}
\nu_t=(C_S\Delta)^2K.
\end{equation}
In mixing-length theories one recovers $\nu_t\propto y$ in the overlap to obtain the log law~\cite{Pope2000}. Here $\nu_{\mathrm{eff}}=\nu_0+s$ is not imposed algebraically but emerges from \eqref{eq:s_evolution}; in the overlap we obtain $s\propto y$ from scale invariance, recovering the mixing-length behavior without postulating it.

\paragraph*{Two-equation models and anisotropy.}
In $k$--$\epsilon$ or $k$--$\omega$ models~\cite{LaunderSpalding1974,Menter1994}, $\nu_t$ is related to independently evolved turbulence quantities. The present framework evolves only one scalar field $s$, but includes a commutator-driven anisotropic correction \eqref{eq:aniso_closure} to capture a minimal rotation-induced misalignment of stress and strain.

\paragraph*{WRLES vs WMLES.}
The WRLES and WMLES limits appear naturally as asymptotic regimes of the same PDE. In the inner region, production is strongly constrained (and vanishes in the strict 2D limit), while in the overlap region scale invariance selects $s\propto y$. Reviews and modern perspectives on wall modeling and WMLES can be found in~\cite{PiomelliBalaras2002ARFM,BosePark2018ARFM}, and unified wall-resolved/wall-modeled strategies include, e.g.,~\cite{DeVanna2021PRF}.

\paragraph*{RANS one-equation baselines.}
For baseline context, one-equation RANS closures such as Spalart--Allmaras~\cite{SpalartAllmaras1992} and hybrid RANS/LES strategies such as DES~\cite{Spalart2009DES} provide useful reference points. The present two-field setting is not intended as a drop-in replacement, but as an alternative geometric organizing principle.

\paragraph*{Complex geometries and immersed boundary methods.}
In complex geometries, wall modeling is often combined with immersed boundary methods and related sharp-interface techniques. While discretization is outside our scope, standard references include Peskin~\cite{Peskin2002IBM} and Mittal \& Iaccarino~\cite{MittalIaccarino2005IBM}. For pressure-gradient boundary-layer context relevant to wall-turbulence modeling, see also~\cite{VinuesaRozier2014AIAA}.

\section{The Geometry of Uncertainty: A Physical Interpretation}
\label{sec:discussion}

\paragraph*{The ontology of $s$ (magnitude).}
The porous-medium morphology term, $\Div(s\grad s)$, supports finite-speed spreading of compactly supported disturbances. In this sense the uncertainty field organizes as a propagating front rather than diffusing instantaneously as in a linear heat equation.

\paragraph*{The emergence of anisotropy (structure).}
The non-linear stress closure $\boldsymbol{\tau}_{\text{aniso}} \propto \frac{s}{K}[\bm{S}, \bm{\Omega}]$ represents a minimal geometric frustration: strain tends to align stress, while rotation induces reorientation. The commutator measures the failure of simultaneous diagonalization of $\bm{S}$ and $\bm{\Omega}$, and thus provides an objective signature of rotation-induced misalignment.

\paragraph*{Causality and the forward cascade.}
Driving production via $\mathcal{I}=\|\bm{S}\boldsymbol{\omega}\|^2$ enforces a kinematic condition: sustained production occurs only where vortex stretching is active, and ceases in strictly two-dimensional limits.

\section{Conclusion}
\label{sec:conclusion}

Building on our previous work~\cite{SanchisAgudoVinuesa2025PRL}, we developed a two-field framework in which the velocity $\ve$ and an intermittency field $s$ evolve in a coupled manner. The effective viscosity is $\nu_{\mathrm{eff}}=\nu_0+s$, and the total stress includes an emergent commutator term $\frac{s}{K}[\bm{S},\bm{\Omega}]$ that breaks pure Boussinesq isotropy.

In a wall-resolved asymptotic limit, constraining production through the vortex-stretching invariant and balancing morphology with relaxation yields $s\sim y^2$ as $y\to 0$. In the overlap layer, high Reynolds number, constant stress and scale invariance yield $s(y)\propto y$ as the unique nontrivial profile, recovering the logarithmic law of the wall.

\begin{acknowledgments}
M.S-A. acknowledges financial support from the EU Doctoral Network MODELAIR. M.S.-A.\ thanks Emelie Saga Stark for numerous discussions and insightful suggestions. Special thanks to Francesco Mario D'Afiero and Eduardo Terres-Caballero for discussions that sharpened the conceptual foundations.
\end{acknowledgments}

\appendix

\section{Derivation of anomalous scaling and K62 intermittency corrections}
\label{app:K62}

This appendix shows how the multiplicative structure of the $s$-dynamics yields log-normal statistics for coarse-grained dissipation and reproduces the Kolmogorov--Obukhov refined similarity hypothesis (K62)~\cite{Kolmogorov1962,Obukhov1962,Frisch1995}.

\subsection{Multiplicative dynamics along Lagrangian trajectories}

Define the material derivative $D_t := \partial_t + \ve\cdot\grad$. In the high-Reynolds-number limit ($\nu_0\ll s$), and on local time scales for which the morphology term is subdominant relative to source/sink terms, Eq.~\eqref{eq:s_evolution} reduces to
\begin{equation}
D_t s \approx q_{\mathrm{prod}} - q_{\mathrm{relax}}.
\label{eq:app_lag_s}
\end{equation}
Using Eqs.~\eqref{eq:q_prod}--\eqref{eq:q_relax}, dropping $\nu_0$ against $s$, and retaining the linear relaxation (dominant in overlap/inertial-range balances), we obtain
\begin{equation}
D_t s \approx s\Big(C_I\,\mathcal{I}(\bm{S},\boldsymbol{\omega}) - K\Big).
\label{eq:app_mult}
\end{equation}
Defining the net amplification rate
\begin{equation}
\xi(t) := C_I(t)\,\mathcal{I}(t) - K(t),
\end{equation}
we get the multiplicative (logarithmic) law
\begin{equation}
D_t \ln s = \xi(t).
\label{eq:app_lns}
\end{equation}
The quadratic saturation term provides nonlinear damping at very large $s$; Eq.~\eqref{eq:app_lns} remains the leading-order driver of multiplicative fluctuations when linear relaxation dominates.

\subsection{Log-normality of $s$}

Integrating Eq.~\eqref{eq:app_lns} along a trajectory from $t=0$ to $t=T$,
\begin{equation}
\ln\frac{s(T)}{s(0)} = \int_{0}^{T}\xi(t')\,\dd t'.
\label{eq:app_int}
\end{equation}
If $\xi(t)$ is stationary with finite correlation time and satisfies a central-limit-type condition for $T$ sufficiently larger than that correlation time, the integral approaches a Gaussian random variable. Hence $\ln s$ is approximately normal and $s$ is approximately log-normally distributed.

\subsection{Coarse-grained dissipation and K62 variance}

For $\nu_0\ll s$,
\begin{equation}
\epsilon(\bm{x},t)=\nu_{\mathrm{eff}}K^2 \approx sK^2.
\end{equation}
Define coarse-grained dissipation over a ball of size $r$,
\begin{equation}
\epsilon_r(\bm{x},t) := \frac{1}{|B_r|}\int_{B_r(\bm{x})}\epsilon(\bm{y},t)\,\dd\bm{y}.
\end{equation}
K62 assumes that the cascade depth scales like $\ln(L/r)$, giving
\begin{equation}
\mathrm{Var}\!\left(\ln \epsilon_r\right) = \mu \ln\!\left(\frac{L}{r}\right),
\label{eq:app_var}
\end{equation}
with intermittency coefficient $\mu$.

Choosing the mean of $\ln\epsilon_r$ to enforce $\langle \epsilon_r\rangle=\epsilon_0$ yields the log-normal moment formula
\begin{equation}
\langle \epsilon_r^{q}\rangle
= \epsilon_0^{q}\left(\frac{L}{r}\right)^{\frac{\mu}{2}q(q-1)}.
\label{eq:app_mom}
\end{equation}

\subsection{Anomalous scaling of structure functions}

Refined similarity gives (in the inertial range)
\begin{equation}
\left\langle |\delta u_r|^{p}\right\rangle
\sim
\left\langle \epsilon_r^{p/3}\right\rangle r^{p/3}.
\label{eq:app_rsh}
\end{equation}
Taking $q=p/3$ in Eq.~\eqref{eq:app_mom} gives
\begin{equation}
\left\langle |\delta u_r|^{p}\right\rangle \sim r^{\zeta_p},
\qquad
\zeta_p = \frac{p}{3} - \frac{\mu}{18}p(p-3),
\label{eq:app_zeta}
\end{equation}
which is the standard K62 log-normal intermittency correction.

\bibliography{twofield}

\end{document}